\documentstyle[epsf,epsfig,aps]{revtex}
\draft
\input epsf
\begin{document}
\twocolumn[\hsize\textwidth\columnwidth\hsize\csname@twocolumnfalse%
\endcsname
\title{Observation of dc voltage on segments of an inhomogeneous superconducting loop}
\author{S. \ V. \ Dubonos, \ V. \ I. \ Kuznetsov, and \ A. \ V. \ Nikulov}

\address{Institute of Microelectronics Technology and High Purity Materials, Russian Academy of Sciences, 142432 Chernogolovka, Moscow District, RUSSIA}

\maketitle
\begin{abstract}
{In order to verify a possibility of a dc voltage predicted on segments of an inhomogeneous superconducting loop the Little-Parks oscillations are investigated on symmetrical and asymmetric Al loops. The amplitude of the voltage oscillations $\Delta V$ measured on segments of symmetrical loop increases with the measuring current $I_{m}$ and $\Delta V = 0$ at $I_{m}= 0$ in accordance with the classical Little-Parks experiment. Whereas the $\Delta V$ measured on segments of asymmetric loop has a maximum value at $I_{m}= 0$. The observation of the dc voltage at $I_{m}= 0$ means that one of the loop segments is a dc power source and others is a load. The dc power can be induced by both thermal fluctuation and a external electric noise. }
\end{abstract}
]
\narrowtext

Recently thought-provoking claims were made about violation of the second
law of thermodynamics in the quantum regime
\cite{Amsterda,Prague,myEprint9}. These claims were made independently by
three research teams from different lands and are concerned to different
branches of knowledge such as quantum thermodynamics, biomolecules,
superconductivity and others. The publication of such sensation statement
has attracted the attention of the scientific press \cite{Schewe} but most
scientists still are not buying  \cite{Weiss} the theoretical arguments
presented in \cite{Amsterda,Prague,myEprint9}.

The present work is devoted to the experimental verification of a
theoretical result \cite{jltp98} according to which a dc voltage can be
observed on segments of a inhomogeneous superconducting loop at $T \approx
T_{c}$ without any external current. The value and sign of this dc voltage
depend in a periodic way on a magnetic flux $\Phi$ within the loop
$\overline{V_{os}}(\Phi/\Phi_{0})$. The work  \cite{jltp98} was provoked by
an experimental observation \cite{Zhilyaev} which is not published up to
now, however. According to the opinion \cite{myEprint9} by one of the
authors of the present and  \cite{jltp98} works the existence of the dc
voltage contradicts to the second law if $\overline{V_{os}}(\Phi/\Phi_{0})$
is induced by the thermal fluctuations in the thermodynamic equilibrium
state.

In order to verify the result \cite{jltp98} we used the mesoscopic Al
structures, one of them is shown on Fig.1. These microstructures are
prepared using an electron lithograph developed on the basis of a
JEOL-840A electron scanning microscop. An electron beam of the lithograph
was controlled by a PC, equipped with a software package for proximity
effect correction "PROXY". The exposition was made at 25 kV and 30 pA. The
resist was developed in MIBK: IPA = 1: 5, followed by the thermal
deposition of a high-purity Al film 60 nm and lift-off in acetone. The
substrates are Si wafers.  The measurements are performed in a standard
helium-4 cryostat allowing us to vary the temperature down to 1.2 K. The
applied magnetic field, which is produced by a superconducting coil, never
exceeded 35 Oe. The voltage variations down to 0.05 $\mu V$ could be
detected.

We have investigated the dependencies of the dc voltage $V$ on the magnetic
flux $\Phi \approx BS$ of some round loops with a diameter 2r = 1, 2 and 4
$\mu m$ and a linewidth w = 0.2 and 0.4 $\mu m$ at the dc measuring current
$I_{m}$ and different temperature closed to $T_{c}$. Here $B$ is the
magnetic induction produced by the coil; $S = \pi r^{2}$ is the area of the
loop. The sheet resistance of the loops was equal approximately $0.5 \
\Omega /\diamond $ at 4.2 K, the resistance ratio $R(300 K)/R(4.2 K)
\approx 2$ and the midpoint of the superconducting resistive transition
$T_{c} \approx  1.24 \ K$. All loops exhibited the anomalous features of
the resistive dependencies on temperature and magnetic field which was
observed on mesoscopic Al structures in some works \cite{repeat,anomal}
before. We assume that these features can be connected with big value of
the Al superconducting coherence length which can exceed a structure size
near $T_{c}$.

\begin{figure}[bhb] \vspace{0.1cm}\hspace{-1.5cm}
\vbox{\hfil\epsfig{figure= 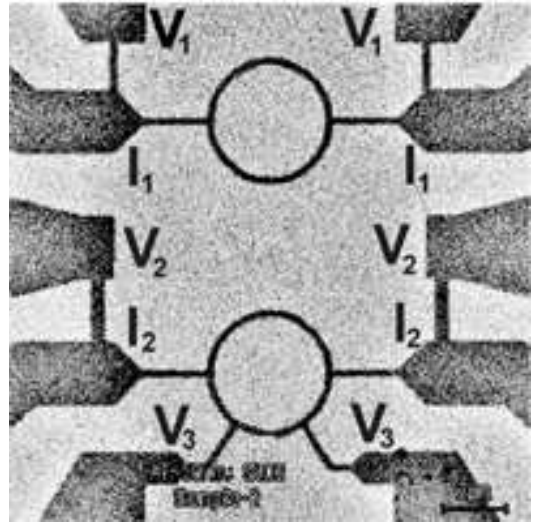,width=7cm,angle=0}\hfil}
\vspace{0.75cm} \caption{An electron micrograph one of the aluminum loop
samples.  $I_{1}$ and $V_{1}$ are the current and potential contacts of
the  symmetrical loop. $I_{2}$ and $V_{2}$ are the current and potential
contacts of the asymmetric loop. $V_{3}$ are the additional potential
contacts of the asymmetric loop.} \label{fig-1} \end{figure}

According to \cite{jltp98} the dc voltage can observed in a inhomogeneous
loop and should not observed in a homogeneous one. In order to investigate
the influence of the heterogeneity of loop segments we made both symmetrical
and asymmetric loops in each investigated structure (see Fig.1). Because of
the additional potential contacts the higher and lower segments of the
lower loop (on Fig.1) can have a different resistance at $T \simeq T_{c}$
when the magnetic flux  $ \Phi$ contained within a loop is not divisible by
the flux quantum $\Phi_{0} = \pi \hbar c/e$, i.e.  $\Phi \neq n\Phi_{0}$,
whereas the one of the higher loop should have the same resistance if any
accidental heterogeneity is absent.

\begin{figure}[bhb] \vspace{0.1cm}\hspace{-1.5cm}
\vbox{\hfil\epsfig{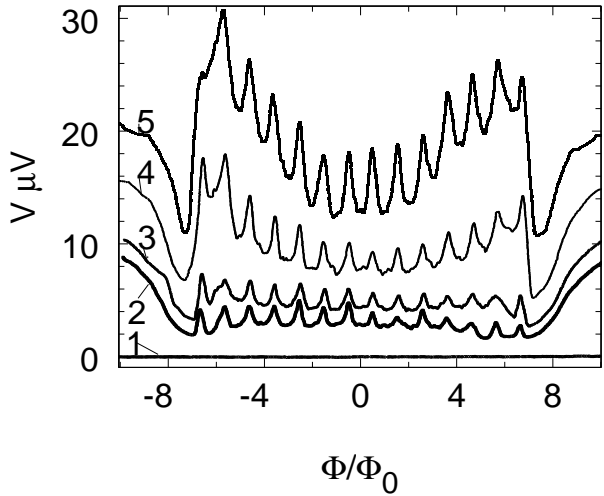}\hfil}
\vspace{0.75cm} \caption{The voltage oscillations measured on the $V_{1}$
contacts of the symmetrical loop with 2r = 4 $\mu m$ and w = 0.2 $\mu m$ at
different $I_{m}$ values between the  $I_{1}$ contacts: 1 - $I_{m} = 0.000
\ \mu A$; 2 - $I_{m} = 1.83 \ \mu A$;  3 - $I_{m} = 2.10 \ \mu A$; 4 -
$I_{m} = 2.66 \ \mu A$; 5 - $I_{m} = 3.01 \ \mu A$. $T = 1.231 K$ is
corresponded to the bottom of the resistive transition } \label{fig-1}
\end{figure}

The voltage oscillations measured on the $V_{1}$ contacts Fig.2 and on the
$V_{2}$ contacts Fig.3 confirm qualitative difference between the symmetrical
and asymmetric loops. In the first case the amplitude $\Delta V$ of the
voltage oscillations increases with the measuring current  $I_{m}$ and the
oscillations are not observed at $I_{m} = 0$. Whereas in the second case
the greatest oscillations are observed at  $I_{m} = 0$ and the  $\Delta V$
value does not increase with the $I_{m}$, Fig.3. Not only the voltage value
but also the sign of the voltage are changed with the magnetic field at
$I_{m} = 0$, Figs.3,4.

In the present work we consider only the region $|\Phi/\Phi_{0}| < 7$ where
the dependencies $V \approx R_{m}(\Phi/\Phi_{0},T/T_{c}(I_{m}))I_{m}$ Fig.2
corresponds to the classical Little-Parks (LP) experiment  \cite{little}.
The anomalous behaviour, the downfall observed before the disappearance of
the oscillation Fig.2, will be considered later. In contrast to the
classical LP experiment no resistance but voltage oscillations are observed
on the asymmetric loop: $V \approx  V_{os}(\Phi/\Phi_{0}) + R_{nos}I_{m}$
Fig.3. The resistance $R_{nos}$ depends faintly on $I_{m}$ and on the
magnetic field at low  $I_{m}$ Fig.3. At a high  $I_{m}$ value the negative
magnetoresistance $R_{nos}$ is observed Fig.3. Such anomaly was observed
also on other our loops and in other works \cite{anomal}.

According to the universally recognized explanation  \cite{tink75} the LP
resistance oscillations are observed  \cite{repeat} because of the fluxoid
quantization \cite{little,tinkham}. The resistance increase at  $\Phi \neq
n\Phi_{0}$ is interpreted as a consequence of the $T_{c}$ decrease at a
non-zero velocity of superconducting pairs $v_{s} \neq 0$:  $\Delta R =-
(dR(T-T_{c})/dT)\Delta T_{c} \propto (dR/dT)v_{s}^{2}$ \cite{tink75}.
Because of the quantization $$\int_{l}dl v_{s} = \frac{\pi \hbar}{m} (n
-\frac{\Phi}{\Phi_{0}}) \eqno{(1)} $$ the $v_{s}$ circulation can not be
equal zero at  $\Phi = BS + LI_{s}\simeq BS \neq n\Phi_{0}$ \cite{tink75}.
At zero measuring current the  $v_{s}$ value is proportional to the
superconducting screening current $v_{s} \propto I_{sc}= s2en_{s}v_{s} =
s2e <n_{s}^{-1}>^{-1}(\pi \hbar/ml) (n -\Phi/\Phi_{0})$. $<n_{s}^{-1}> =
l^{-1}\int_{l}dl n_{s}^{-1}$ is used because the superconducting current
$I_{s} = sj_{s} =s 2en_{s}v_{s}$ should be constant along the loop in the
stationary state. At $I_{m} \neq 0$ and $<n_{s}^{-1}>^{-1} \neq 0$ in one
of the loop segments  $|v_{s}| \propto |I_{m}/2 + I_{sc}|$ and in the other
one $|v_{s}| \propto |I_{m}/2 - I_{sc}|$.

\begin{figure}[bhb]
\vspace{0.1cm}\hspace{-1.5cm}
\vbox{\hfil\epsfig{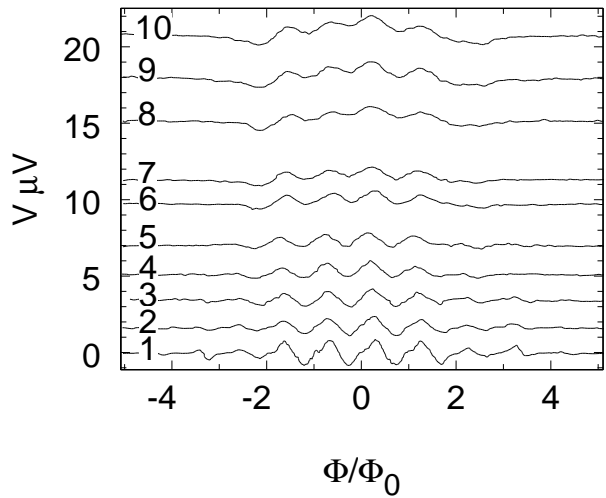}\hfil}
\vspace{0.75cm} \caption{The voltage oscillation measured on the $V_{2}$
contacts of the asymmetric loop with 2r = 4 $\mu m$ and w = 0.4 $\mu m$ at
different value of the measuring current between the  $I_{2}$ contacts: 1 -
$I_{m} = 0.000 \ \mu A$; 2 - $I_{m} = 0.29 \ \mu A$;  3 - $I_{m} = 0.65 \
\mu A$; 4 - $I_{m} = 0.93 \ \mu A$;  5 - $I_{m} = 1.29 \ \mu A$;  6 -
$I_{m} =1.79 \ \mu A$;  7 - $I_{m} = 2.06 \ \mu A$;  8 - $I_{m} = 2.82 \
\mu A$;  9 - $I_{m} = 3.34 \ \mu A$; 10 - $I_{m} =3.85 \ \mu A$.  $T =
1.231 K$ is corresponded to the bottom of the resistive transition }
\label{fig-1} \end{figure}

Because  $I_{sc} = 0$, $R_{hs} \neq 0$ or/and $R_{ls} \neq 0$ at
$<n_{s}^{-1}>^{-1} = 0$ when any loop segment in the normal state, i.e. the
density of superconducting pairs $n_{s} = 0$, and  $R_{hs} = 0$,  $R_{ls} =
0$ at $<n_{s}^{-1}>^{-1} \neq 0$ when the whole loop is in the
superconducting state, i.e. $n_{s} \neq 0$ along the whole loop, the LP
oscillations are observed only near $T_{c}$ where the switching between the
states with $<n_{s}^{-1}>^{-1} \neq 0$ and $<n_{s}^{-1}>^{-1} = 0$ take
place and the $I_{sc}$, $R_{hs}$, $R_{ls}$ values change in time. Here
$R_{hs}$ and $R_{ls}$ are the resistance of the higher and  lower  segments
in a stationary state. The resistance $R_{m} = V/I_{m}$ and the voltage $V$
measured at the LP experiment are the average in time values: $V =
\overline{V(t)} = t_{long}^{-1} \int_{t_{long}}dt V(t)$; $R_{m} \approx
\overline{(1/R_{hs} + 1/R_{ls})^{-1}} = \sum P(R_{hs},R_{ls})(1/R_{hs} +
1/R_{ls})^{-1}$. Where $P(R_{hs},R_{ls})$ is the probability of the states
with non-zero $R_{hs}$ and  $R_{ls}$ values.

According to \cite{tink75} not only the average $\overline{I_{sc}^{2}} =
t_{long}^{-1} \int_{t_{long}}dt I_{sc}^{2}$ but also $s\overline{j_{sc}} =
\overline{I_{sc}} = t_{long}^{-1} \int_{t_{long}}dt I_{sc} \approx
s2e\overline{<n_{s}^{-1}>^{-1}}(\pi \hbar/ml)\overline{(n -\Phi/\Phi_{0})}$
is not equal zero at $ \Phi \neq n\Phi_{0}$ and $ \Phi \neq (n+0.5)
\Phi_{0}$. The theoretical dependence  $\Delta  T_{c} \propto -(n
-\Phi/\Phi_{0})_{min}^{2}$, where $n$ is corresponded  to minimum possible
value $v_{s}^{2} \propto (n -\Phi/\Phi_{0})^{2}$ \cite{tink75} describes
enough well the experimental data (see for example Fig.4 in
\cite{repeat}).  Therefore $\overline{(n -\Phi/\Phi_{0})} \approx  (n
-\Phi/\Phi_{0})_{min}$ when $\Phi$ is not close to  $(n+0.5) \Phi_{0}$.
$\overline{I_{sc}} = 0$ at  $ \Phi = (n+0.5) \Phi_{0}$ because the
permitted states with opposite $v_{s}$ direction have the same $v_{s}^{2}$
value. Thus, the LP experiment is evidence of the persistent screening
current $I_{p.c.} = \overline{I_{sc}}$ flows along the loop at a constant
magnetic flux, $ \Phi \neq n\Phi_{0}$ and $ \Phi \neq (n+0.5) \Phi_{0}$,
and $R_{l} \neq 0$.

It is enough obvious from the analogy with a conventional loop that the
potential difference $\overline{V_{sc}} = (<\overline{\rho}>_{l_{s}} -
<\overline{\rho}>_{l})l_{s}\overline{j_{sc}}$ can be observed on a segment
$l_{s}$ of an inhomogeneous loop at $\overline{j_{sc}}$ if the average
resistivity along the segment $<\overline{\rho}>_{l_{s}} = \int_{l_{s}} dl
\overline{\rho}/l_{s}$ differs from the one along the loop
$<\overline{\rho}>_{l} = \int_{l} dl \overline{\rho}/l$.  Because the
$\overline{I_{sc}} (\Phi/\Phi_{0})$ oscillations take place both in the
symmetrical and asymmetric loops the absence of the voltage oscillations at
$I_{m} = 0$ on the contacts $V_{1}$ Fig.2 and the observation on $V_{2}$
Figs.3,4 mean that $\overline{R_{hs}} = \overline{R_{ls}}$ in the first
case and $\overline{R_{hs}} \neq \overline{R_{ls}}$ in the second case. The
later can be if the critical temperature of the higher and lower segments
are different: if $T_{ch}(\Phi) \neq T_{cl}(\Phi)$ then
$\overline{R_{hs}}(T - T_{ch}) \neq \overline{R_{ls}}(T - T_{cl})$ at $T
\approx T_{ch},T_{cl}$.

Consequently, the voltage oscillations at $I_{m} = 0$ Fig.3,4 can be caused
by the superconducting screening current, as well as the LP oscillations
Fig.2. The comparison of the experimental data for symmetrical and asymmetric
loops confirms this supposition. Our investigations have shown that the
voltage oscillations as well as the LP oscillations were observed only in
the temperatures corresponded to the resistive transition.  Both
oscillations have the same period. The magnetic field regions, where they
are observed, are also closed.

The oscillations on Fig.2 are observed in more wide magnetic field region
than on Figs.3,4 because the width of the wire defining the loop in the
first case w = 0.2 $\mu m$ is smaller than in the second case w = 0.4 $\mu
m$. In any real case only some oscillations are observed because a high
magnetic field breaks down the superconductivity, i.e $I_{sc}$, in the wire
defining the loop and the contact grounds. According to (1) $v_{s} = (\pi
\hbar /m) Br/2$ along the loop and $v_{s} = (\pi \hbar /m) Bw/2$ along the
boundaries of the wire at $n=0$. Therefore a limited number of oscillations
$\propto 2r/w$ are observed.  The wide contact grounds, with the width
$\approx 2 \ \mu m$ (see Fig.1), have also an influence on the oscillation
number.

According to the analogy with a conventional loop the voltage measured
between the $V_{2}$ contacts  $\overline{V_{sc}} = 0.5(\overline{R_{hs}} -
\overline{R_{ls}})\overline{I_{sc}}$ and consequently the voltage
oscillations with the amplitude $\Delta V \approx 1 \ \mu V$ observed on
Fig.4 can be induced by $\overline{I_{sc}}$ oscillations with  $\Delta
\overline{I_{sc}} \geq 0.4 \ \mu A$ because
$\overline{R_{hs}},\overline{R_{ls}} \leq R_{ln}/2 = 5 \ \Omega$. The
screening current $\overline{|I_{sc}|}$ inducing the $R_{m}$ oscillations
Fig.2 can be evaluate from the experimental data if the
$dR_{m}/d\overline{|I_{sc}|}$ value is known. Although $|I_{m}/2 + I_{sc}|
> |I_{m}/2|$ in one of the segments and $|I_{m}/2 - I_{sc}| < |I_{m}/2|$ in
the other one at $|I_{m}/2| > |I_{sc}|$ $dR_{m}/d\overline{|I_{sc}|} > 0$
and the LP oscillations are observed at both small and large $I_{m}$ (see
Fig.2 and \cite{repeat}) because  $I_{sc}$, as well  $I_{m}$, decreases the
probability of superconducting state $<n_{s}^{-1}>^{-1} \neq 0$ and
consequently increases the $P(R_{hs} \neq 0,R_{ls} \neq 0)$. If one assumes
that the $dR_{m}/d|I_{m}|$ and $dR_{m}/d\overline{|I_{sc}|}$ are closed in
order of value then according to the data presented on Fig.2
$\overline{|I_{sc}|} \approx 0.4 \ \mu A$ at $\Phi = (n+0.5) \Phi_{0}$.

\begin{figure}[bhb]
\vspace{0.1cm}\hspace{-1.5cm}
\vbox{\hfil\epsfig{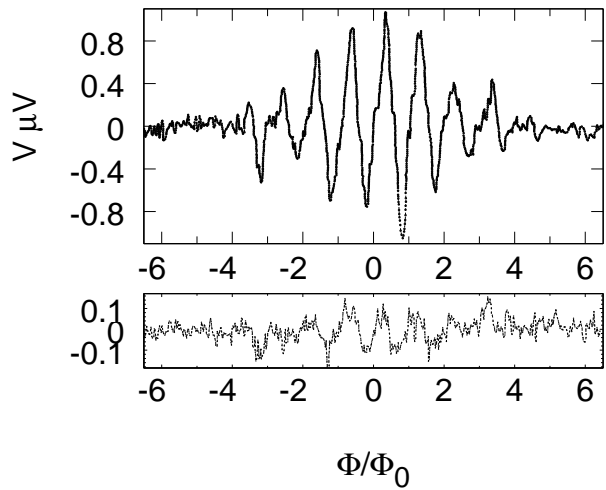}\hfil}
\vspace{0.75cm}
\caption{Oscillation of the voltage measured on the $V_{2}$ contacts (upper
curve) and on the $V_{3}$ contacts  (lower curve) of the asymmetric loop
with 2r = 4 $\mu m$ and w = 0.4 $\mu m$. $I_{m} = 0$. $T = 1.231 K$
corresponded to the bottom of the resistive transition. } \label{fig-1}
\end{figure}

Thus,  the analogy with a conventional loop and  the classical LP
experiment Fig.2 explain enough well the voltage oscillations observed at
$I_{m} = 0$ Figs.3,4. But in contrast to the case of the conventional loop
when the current $I_{sc} = R_{l}(-1/c)d\Phi/dt$ and the electric field $E =
-\bigtriangledown V - (1/c)dA/dt =  -\bigtriangledown V - (1/cl)d\Phi/dt$
have the same direction in both segment in our case $\overline{d\Phi/dt} =
0$ and consequently the average current  $\overline{I_{sc}}$ and the
average electric field $\overline{E} = -\overline{\bigtriangledown V}$
should have opposite directions in one of the segments because $\int_{l}dl
\bigtriangledown V \equiv 0$. This means that one of the loop segments is a
dc power source $W = \overline{V_{os}I_{sc}}$ and others is a load.

The existence of the dc power contradicts to some habitual knowledge if $W$
is not induced by a temperature difference $\Delta T$. The voltage
oscillations Fig.4 can be explain by an accidental temperature difference
$V_{os} = S_{th}\Delta T$ only if the thermopower $S_{th}$ is oscillated
and its sign is switched together with the $\overline{I_{sc}}$. The
thermopower oscillations are observed in some Andreev interferometer
\cite{thermopo} but its value is very small in order to explain the
voltage oscillations observed in our work.

Because the voltage and LP oscillations are observed in the same region it
is naturally to explain the observation of the dc power as a direct
consequence of the contradiction of the LP experiment with the Ohm's law
$R_{l}I_{sc} = \int_{l}dl E = -(1/c)d\Phi/dt$ and other fundamental laws
\cite{QuaForce}.  The existence of $\overline{I_{sc}} \neq 0$ at
$\overline{R_{l}} \neq 0$ and $\overline{d\Phi/dt} = 0$ is explained
\cite{QuaForce} by the change of the momentum circulation of
superconducting pairs from $\int_{l}dl p = \int_{l}dl (2mv_{s} + (2e/c)A) =
(2e/c)\Phi$ at $<n_{s}^{-1}>^{-1} = 0$ to $\int_{l}dl p = n2\pi \hbar$ at
$<n_{s}^{-1}>^{-1} \neq 0$ at the closing of superconducting state, when
its connectivity changes.

These momentum changes because of the quantization  $n2\pi \hbar -
(2e/c)\Phi = 2\pi \hbar(n - \Phi/\Phi_{0})$ take the place of the Faraday's
voltage $- (1/c) d\Phi/dt$. The force maintaining the persistent current,
as well as the Faraday's electric field $ - (1/cl) d\Phi/dt$, should be
uniform along the loop because  the momentum change on the unit volume
$\Delta P \propto  j_{s}$ \cite{QuaForce}. This warrants the analogy with a
conventional loop used above.

At a enough low frequency, when the switching takes place between the
stationary states $$ \overline{V_{os}} =(
\frac{<\overline{\rho}>_{l_{s}}}{<\overline{\rho}>_{l}} - 1)
\frac{l_{s}}{l} \frac{\pi \hbar}{e}\overline{ (n
-\frac{\Phi}{\Phi_{0}})}\omega \eqno{(2)} $$ on a $l_{s}$ segment. $\omega
= N_{sw}/t_{long}$ is the average frequency of a switching between the
superconducting state with different connectivity; $N_{sw}$ is the number
of switching for $t_{long}$. The amplitude of the oscillations $\Delta
\overline{V_{os}} \leq 0.25 (\pi \hbar/e)\omega$ at $l_{s} = l/2$. $\pi
\hbar/e = 2.07 \ \mu V/GHz$ is equal to the ratio of the voltage and the
frequency in the Josephson effect \cite{Barone}. Consequently, according to
(2) the oscillations Fig.4 with $\Delta V \approx 1 \ \mu V$ can be
observed if $\omega \geq 2 \ GHz$.

The $\Delta  \overline{V_{os}}$ increases more slowly with the frequency
than (2) if $\omega > 1/\tau _{rel}$. Where $\tau _{rel}$ is any relaxation
time in stationary states, which can be equal the relaxation time of
superconducting fluctuations $\tau _{fl}$ \cite{tink75} or  the decay time
of the screening current $\tau _{R}$. $1/\tau _{fl}\approx  2 \ GHz$ in
order of value because  $\tau _{fl} = \pi \hbar /8k_{B}(T - T_{c})$ in the
linear approximation region above $T_{c}$ \cite{tink75} and the width of
the critical region of our loops $ \Delta T_{r.t.} \approx 0.02 K$.
$1/\tau _{R} \approx (2e^{2}/m)n_{s}\rho _{n} \approx e\rho
_{n}\overline{j_{sc}}/mv_{s} \approx 10 \ GHz$  in order of value. Here the
value  $\overline{|j_{sc}|} = \overline{|I_{sc}|}/s \approx 2 \ 10^{7}  \
A/m^{2}$ found above, $|v_{s}| = \pi \hbar /2ml \approx 30 \ m/s$ for $|n -
\Phi/\Phi_{0}| = 0.5$ and the $\rho _{n}$ value of Al were used.
Consequently,  the voltage oscillations observed in our work Fig.3,4 can be
induced by a switching between the superconducting state with different
connectivity.

This switching can be induced by both the thermal fluctuations and an
external electric noise. A high-frequency noise with $I_{noise} \geq sj_{c}
= (c^{2}sk_{B}T/2\pi \lambda ^{2}\xi )^{1/2}$ increases the probability
$P_{sw}$ of the switching in the normal state.  $P_{sw} \propto
exp(-sl_{s}f_{sup}/k_{B}T)$ at $l_{s} \geq \xi $, where $f_{sup} = (2\pi
/c^{2})\lambda ^{2}j_{c}^{2}$ is the energy density of the transition  in
the normal state \cite{tink75}; $\xi$ is the superconducting coherence
length; $\lambda$ is the London penetration depth. For the loops used in
our work  $(c^{2}sk_{B}T/2\pi \lambda ^{2}\xi )^{1/2} \approx (\Delta
T_{r.t.}/T_{c})^{3/4} \ 10^{-5} \ A \approx  0.5 \ \mu A$. We can not
guarantee that $I_{noise} \ll 0.5 \ \mu A$. Moreover we observed an
influence of an external electric noise on the $\overline{V_{os}}$ value.
Therefore we can not state that the voltage oscillations observed in our
work at $I_{m} = 0$ are induced in the equilibrium state although the power
$W = \overline{V_{os}I_{sc}} \approx 2 \ 10^{-13} Wt $ does not exceed the
limit value $k_{B}T/\hbar \approx 10^{-12}$ \cite{myEprint9,QuaForce} which
can be induced by the thermal fluctuations.

In conclusion, we have observed voltage oscillations measured on segments
of an inhomogeneous loop at zero external direct current in the same region
where the Little-Parks oscillations are observed. This voltage can be
induced by both thermal fluctuation and an external electric noise.

We acknowledge useful discussions with V.A.Tulin.

\end{document}